\documentclass[aps,prb,a4paper,preprint,showpacs]{revtex4}
\usepackage{graphicx}
\usepackage{amsmath}
\usepackage{amsfonts}
\usepackage{amssymb}
\usepackage{color}
\usepackage{amsbsy}
\usepackage{bm}
\usepackage{epstopdf}
\usepackage{epsfig}

\begin{document}
\title{Topological phase transitions driven by polarity change and next-nearest-neighbor hopping in skyrmion crystals}
\author{ Jianhua Gong and Rui Zhu\renewcommand{\thefootnote}{*}\footnote{Corresponding author.
Electronic address:
rzhu@scut.edu.cn} }
\address{School of Physics and Optoelectronics, South China University of Technology,
Guangzhou 510641, People's Republic of China   }

\begin{abstract}

By considering the bulk-edge correspondence of the skyrmion crystal this work focused on the relation of the topological properties to the polarity ($Q_{\rm{sk}}$), next-nearest-neighbor hopping ($t'$), and exchange coupling ($J$). We found that by continually increasing polarity of the skyrmions from 1 to 2, the monopole-lattice phase and the dipole-lattice phase both span a wide range in the parameter space demonstrating topological robustness. While the next-nearest-neighbor-hopping of the electrons coexists with the nearest-neighbor hopping the flux pattern of the emergent magnetic field traversing the two-dimensional SkX is reshaped giving rise to different topological Hall effect. By increasing $t'$ from 0 to $t$ with the polarity fixed to be 1, we found Chern numbers of the electronic bands change from unity to indefinite values demonstrating a phase transition. We also considered the effect of finite $J$ and found the results obtained in the adiabatic limit hold to around $J=t$.

\end{abstract}

\maketitle
\section{Introduction}

The skyrmion crystal (SkX) is a two-dimensional (2D) lattice structure formed by periodically-aligned magnetic skyrmion vortices\cite{NagaosaNatNano2013, MuhlbauerScience2009, YingSuPRRes2020, HayamiPRB2019, OzawaPRL2017}. The spin field ${\bf{n}}({\bf{r}})$ of a single skyrmion has a nontrivial topological number $Q_{\rm{sk}}$ equal to the surface
integral of the solid angle of ${\bf{n}}({\bf{r}})$. While conducting electrons pass the spin field of the SkX, their spins tend to align to the direction of the local magnetization. Because tunneling of electrons can only occur between identical spin states, such an alignment gives rise to a site-dependent hopping integral. In the SkX, each skyrmion vortex can be viewed as a giant unit cell of multiple atoms with sublattice-dependent spin polarizations. The effective hopping integral can be expressed by the uniform magnitude multiplied by a \emph{sublattice-dependent} phase factor. By comparison with the Peierls phase induced by a real magnetic field, conducting electrons moving in the SkX gain a sublattice-dependent ``emergent" magnetic field determined by the local magnetization\cite{NagaosaNatNano2013, HofstadterPRB1976}. The emergent magnetic field perpendicular to the SkX plane can be obtained from ${\bf{n}}({\bf{r}})$ by ${\textstyle{1 \over 2}}{\bf{n}} \cdot ({\textstyle{{\partial {\bf{n}}} \over {\partial x}}} \times {\textstyle{{\partial {\bf{n}}} \over {\partial y}}})$. The topological number of s single skyrmion is the emergent magnetic flux traversing the area of the skyrmion divided by $2 \pi$. The emergent magnetic field gives rise to gapped Landau-level-like bands, which results in quantized topological Hall conductivity and has been confirmed by experiment\cite{NeubauerPRL2009, SchulzNatPhys2012}.

By comparing the emergent magnetic field with the real magnetic field, conducting electrons moving in a SkX background are analogous to those moving in the magnetic field. Considering the nontrivial topology of the band structure, conducting electrons moving in a SkX background are analogous to those moving in various 2D multi-sublattice atomic crystals, such as graphene, Lieb lattice, Dice lattice, and etc. Because the next-nearest-neighbor (nnn) hopping reshapes the flux pattern of the emergent magnetic field and topology of the band structure beyond the nearest-neighbor (nn) hopping, it plays an important role in the transport properties of the conducting electrons. Recently it was found that the nnn hopping drives topological phase transitions in Dirac systems such as the Lieb, kagome, and $T_3$ lattices\cite{BeugelingPRB2012} and non-Dirac diatomic square lattice\cite{OstahiePRB2018} in the tight-binding description. It was also found that third-neighbor-hopping incurs formation of the SkX with a high topological number $Q_{\rm{sk}}=2$ in the Kondo lattice model of itinerant electrons\cite{HayamiPRB2019}. This model provides a candidate mechanism of the SkX formation in inversion-symmetric magnets, which is different from the inversion-asymmetric Dzyaloshinskii-Moriya interaction. Because further-neighbor hopping is important both in the SkX formation and topological transport properties, in this work we investigated the effect of the nnn hopping on the topological phases of the SkX by considering the Chern number of the bulk band and edge states of the nanoribbon geometry. We found that although the nnn hopping changes the shape of the bulk and nanoribbon bands, the Chern number and edge-state number retains unchanged up to a large value of the nnn hopping integral and enters into a non-topological state afterwards, demonstrating a strong topological phase.

Concerning the symmetry of the skyrmion, the spin field of a single skyrmion in the SkX can be expressed by ${\bf{n}}({\bf{r}}) = [\cos \Phi (\varphi )\sin \Theta (r),\sin \Phi (\varphi )\sin \Theta (r),\cos \Theta (r)]$, with $r$ and $\varphi$ being
the polar coordinates in the real space counting from the center of the skyrmion, and $\Theta$ and $\Phi$ being the polar
and azimuthal angles of the atomic magnetization. The topological number $Q_{\rm{sk}}$ of a single skyrmion can be obtained as ${Q_{{\rm{sk}}}} = {\textstyle{1 \over {4\pi }}}[\cos \Theta (r)]_{r = 0}^{r = \infty }[\Phi (\varphi )]_{\varphi  = 0}^{\varphi  = 2\pi }$. When the center spin points downward and the edge spin points upward, which can be assumed\cite{HamamotoPRB2015} to be $\Theta (r)={\pi}(1-r/{\lambda})$ for $r<{\lambda}$ and $\Theta (r)=0$ for $r>{\lambda}$, we can have the vorticity of the skyrmion $m = [\cos \Theta (r)]_{r = 0}^{r = \infty } = 1$. When the skyrmion whirls in
the pattern of $\Phi (\varphi ) = \xi \varphi  + \gamma $, the topological number is explicitly
expressed by ${Q_{{\rm{sk}}}} = m\xi $. The polarity of the skyrmion is defined by $\xi$: $\xi =1$ is a monopole, $\xi =2$ is a dipole (See Fig. 2). $\gamma$ determines the helicity of the skyrmion, which does not affect the topological transport properties of the SkX. Recently, numerical simulation demonstrates that high-topological-number SkX and isolated skyrmion can be created in various systems by different mechanisms\cite{HayamiPRB2019, OzawaPRL2017, XichaoZhangPRB2016, GalkinaPRB2009}. While $m \equiv 1$ in the usual case, the topological number is equal to the polarity of the skyrmion $Q_{\rm{sk}}={\xi}$and the high-topological-number SkX with $Q_{\rm{sk}}=2$ is no other than a dipole lattice [the skyrmion profile is shown in Fig. 2 (e)]. The topological properties of the monopole and dipole are sharply distinct. We recently found that different from the
conventional monopole-lattice SkX, the Hall conductivity quantization number of the dipole-lattice SkX increases by 0, 1, 0, 1, $\cdots$ consecutively when the elevating Fermi energy crosses each band\cite{RuiZhuAIPAdv2019}. Both the skyrmion profile and the topological Hall conductivity show that the monopole lattice and dipole lattice are two distinct topological phases, with the polarity value their distinguishing index. While the conventional SkX ($Q_{\rm{sk}}=1$) and high-topological SkX with $Q_{\rm{sk}}=2$ are definite topological states and various promising applications of the skyrmion by exploiting its topological protection have been proposed\cite{JonietzScience2010, XZhangSR2015, YZhouNatCommun2014, RommingScience2013}, it also wonders us whether the topological state is robust against the change of the polarity. By calculating the bulk Chern number and nanoribbon edge states of the SkX with its polarity varying continuously from monopole to dipole, we found the phase transition between the two states and that the topological state survives to more than $20\% $ deviation of either of the monopole and dipole states, confirming the topological robustness.

The conducting electrons moving in the SkX can be described by the $t-J$ model [see Eq. (\ref{Hamiltonian})]. While the magnitude of the Hund's coupling $J$ is much larger than the hopping energy $t$ and $t'$ of the conducting electrons, the electron spin adiabatically aligns with local magnetization of the SkX and zero-temperature topological Hall conductivity is well quantized measuring the accumulated Chern number below the Fermi energy. While the Hund's coupling becomes weak, the spin degree of the conducting electrons is partly set free, thus weakening its topological nature gained from the spin field of the SkX. Recently, it is found that with weaker Hund's coupling, the topological Hall conductivity becomes unquantized and varies with the strength of the Hund's exchange\cite{DenisovSR2017, DenisovPRB2018, NakazawaJPSJ2018, NakazawaPRB2019}. In this case, we must extend our investigation of the topological phase transitions driven by the polarity and nnn hopping from the condition of strong Hund's coupling to weak Hund's coupling, which constitutes the last part of our investigation.

The other parts of the letter are organized as follows. In Sec. II, we provide model and technique to investigate the topological properties of the conducting electrons in the SkX. In Sec. III A, we discuss the topological phase transition driven by the polarity change. In Sec. III B, we discuss the topological phase transition driven by the nnn hopping. In Sec. III C, we discuss effect of the magnitude of Hund's coupling on the topological phase transitions. Finally, Sec. IV is conclusion, experimental relevance, and outlook for further studies.

\section{Model and formalism}

By taking into account the nnn hopping, the free-electron system coupled with the background spin texture of the SkX can be described by the extended double-exchange model\cite{HamamotoPRB2015}
\begin{equation}
H = t\sum\limits_{ < i,j > } {c_i^\dag {c_j}}  + t'\sum\limits_{ \ll i,j \gg } {c_i^\dag {c_j}}  - J\sum\limits_i {{{\bf{n}}_i}c_i^\dag {\bf{\sigma }}{c_i}} ,
\label{Hamiltonian}
\end{equation}
where ${c_i} = {({c_{i \uparrow }},{c_{i \downarrow }})^{\rm{T}}}$ is the two-component annihilation operator at
the $i$ site and ${c_i^\dag }$ is its creation counterpart. $t$ and $t'$ are the hopping integrals
between nn and nnn sites, respectively. $J$ is the strength of the Hund's coupling between the electron spin and the spin texture of the SkX. ${\bf{\sigma }}$ denotes the Pauli matrix vector. As shown in Fig. 1, the SkX has two lattice structures, i.e., the atomic lattice and the skyrmion lattice. Therefore, we can take each skyrmion to be a giant unit cell and rewrite Hamiltonian (\ref{Hamiltonian}) in the multi-sublattice manner
\begin{equation}
\begin{array}{l}
H = t\sum\limits_{n,m, < s,s' > } {c_{nm}^{s\dag }c_{nm}^{s'}}  + t'\sum\limits_{n,m, \ll s,s' \gg } {c_{nm}^{s\dag }c_{nm}^{s'}}  - J\sum\limits_{n,m,s} {{{\bf{n}}_s}c_{nm}^{s\dag }{\bf{\sigma }}c_{nm}^s} ,\\
 + \left[ {t\sum\limits_{n,m, < s,s' > } {c_{n,m + 1}^{s\dag }c_{nm}^{s'}}  + t\sum\limits_{n,m, < s,s' > } {c_{n + 1,m}^{s\dag }c_{nm}^{s'}}  + {\rm{H}}{\rm{.c}}{\rm{.}}} \right]\\
 + \left[ {t'\sum\limits_{n,m, \ll s,s' \gg } {c_{n,m + 1}^{s\dag }c_{nm}^{s'}}  + t'\sum\limits_{n,m, \ll s,s' \gg } {c_{n + 1,m}^{s\dag }c_{nm}^{s'}}  + {\rm{H}}{\rm{.c}}{\rm{.}}} \right].
\end{array}
\label{HInSublatticeManner}
\end{equation}
We consider each skyrmion unit cell contains $5 \times 5 =25$ atoms. Because both nn and nnn hopping can occur either between different sublattices from the same unit cell or between different sublattices from neighboring unit cells, for the convenience of Fourier transformation later on, we separate intra-unit-cell hopping and inter-unit-cell hopping into different terms in Eq. (\ref{HInSublatticeManner}): the first two terms are intra-unit-cell hopping and the last two terms are inter-unit-cell hopping. $s$ and $s'$ are sublattice labels (circled A to Y in Fig. 1). The spin field is sublattice-dependent instead of site-dependent, which is expressed in the third term of Eq (\ref{HInSublatticeManner}).

In the strong-coupling limit, i.e., $J \gg t,t'$, the spin of the conducting electron is forced to align parallel to the local atomic magnetization. Because there is no other spin-flipping mechanism, the effective hopping strength is determined by the spin overlap between neighboring sites $i$ and $j$, or, equivalently, between sublattices $s$ and $s'$, which can be obtained by $\left\langle {{\chi _s}} \right|\left. {{\chi _{s'}}} \right\rangle $ with $\left| {{\chi _s}} \right\rangle $ being the spin eigenstate of ${{{\bf{n}}_s}}$. Using spherical coordinates in the spin space, we can obtain
\begin{equation}
\left| {{\chi _s}} \right\rangle  = {\left( {\cos \frac{{{\Theta _s}}}{2},{e^{i{\Phi _s}}}\sin \frac{{{\Theta _s}}}{2}} \right)^{\rm{T}}},
\end{equation}
and
\begin{equation}
\left\langle {{\chi _s}} \right|\left. {{\chi _{s'}}} \right\rangle  = \cos \frac{{{\Theta _s}}}{2}\cos \frac{{{\Theta _{s'}}}}{2} + \sin \frac{{{\Theta _s}}}{2}\sin \frac{{{\Theta _{s'}}}}{2}{e^{ - i({\Phi _s} - {\Phi _{s'}})}}.
\end{equation}
In this limit, we arrive at the effective ``tight-binding" model
\begin{equation}
\begin{array}{l}
{H_{{\rm{eff}}}} = \sum\limits_{n,m, < s,s' > } {t_{{\rm{eff}}}^{s,s'}d_{nm}^{s\dag }d_{nm}^{s'}}  + \sum\limits_{n,m, \ll s,s' \gg } {t{'_{{\rm{eff}},s,s'}}d_{nm}^{s\dag }d_{nm}^{s'}} \\
 + \left[ {\sum\limits_{n,m, < s,s' > } {t_{{\rm{eff}}}^{s,s'}d_{n,m + 1}^{s\dag }d_{nm}^{s'}}  + \sum\limits_{n,m, < s,s' > } {t_{{\rm{eff}}}^{s,s'}d_{n + 1,m}^{s\dag }d_{nm}^{s'}}  + {\rm{H}}{\rm{.c}}{\rm{.}}} \right]\\
 + \left[ {\sum\limits_{n,m, \ll s,s' \gg } {t{'_{{\rm{eff}},s,s'}}d_{n,m + 1}^{s\dag }d_{nm}^{s'}}  + \sum\limits_{n,m, \ll s,s' \gg } {t{'_{{\rm{eff}},s,s'}}d_{n + 1,m}^{s\dag }d_{nm}^{s'}}  + {\rm{H}}{\rm{.c}}{\rm{.}}} \right],
\end{array}
\label{EffectiveHamiltonian}
\end{equation}
with $t_{{\rm{eff}}}^{s,s'}{\rm{ = }}t\left\langle {{\chi _s}} \right|\left. {{\chi _{s'}}} \right\rangle $ and $t{'_{{\rm{eff}},s,s'}}{\rm{ = }}t'\left\langle {{\chi _s}} \right|\left. {{\chi _{s'}}} \right\rangle $. Here, $d_{nm}^{s\dag }$ ($d_{nm}^{s }$) is the spinless creation (annihilation) operator at the $nm$ skyrmion unit cell and the $s$ sublattice. By Fourier transformation of Eq. (\ref{EffectiveHamiltonian}), we can obtain
\begin{equation}
\begin{array}{l}
{H_{{\rm{eff}}}} = \sum\limits_{{\bf{k}}, < s,s' > } {t_{{\rm{eff}}}^{s,s'}d_{\bf{k}}^{s\dag }d_{\bf{k}}^{s'}}  + \sum\limits_{{\bf{k}}, \ll s,s' \gg } {t{'_{{\rm{eff}},s,s'}}d_{\bf{k}}^{s\dag }d_{\bf{k}}^{s'}} \\
 + \left[ {\sum\limits_{{\bf{k}}, < s,s' > } {t_{{\rm{eff}}}^{s,s'}d_{\bf{k}}^{s\dag }d_{\bf{k}}^{s'}{e^{ - 2i{k_y}\lambda }}}  + \sum\limits_{{\bf{k}}, < s,s' > } {t_{{\rm{eff}}}^{s,s'}d_{\bf{k}}^{s\dag }d_{\bf{k}}^{s'}{e^{ - 2i{k_x}\lambda }}}  + {\rm{H}}{\rm{.c}}{\rm{.}}} \right]\\
 + \left[ {\sum\limits_{{\bf{k}}, \ll s,s' \gg } {t{'_{{\rm{eff}},s,s'}}d_{\bf{k}}^{s\dag }d_{\bf{k}}^{s'}{e^{ - 2i{k_y}\lambda }}}  + \sum\limits_{{\bf{k}}, \ll s,s' \gg } {t{'_{{\rm{eff}},s,s'}}d_{\bf{k}}^{s\dag }d_{\bf{k}}^{s'}{e^{ - 2i{k_x}\lambda }}}  + {\rm{H}}{\rm{.c}}{\rm{.}}} \right].
\end{array}
\label{BulkHInKSpace}
\end{equation}
This Hamiltonian is quadratic and diagonal in the ${\bf{k}}$-space and a $25 \times 25$ matrix in the sublattice space. By exact diagonalization of it at each point in the momentum space, the band structure $E_{n{\bf{k}}}$ and the electronic states $\left| {n,{\bf{k}}} \right\rangle $ can be obtained with $n = 1,2, \cdots 25$.
The Chern number of the $n$-\emph{th} band can be calculated by\cite{HamamotoPRB2015, ThoulessPRL1982}
\begin{equation}
{{\cal C}_n} = \frac{{ - i}}{{2\pi }}\int_\Omega  {\sum\limits_{m \ne n} {\frac{{\left\langle {n{\bf{k}}} \right|\frac{{\partial H}}{{\partial {k_x}}}\left| {m{\bf{k}}} \right\rangle \left\langle {m{\bf{k}}} \right|\frac{{\partial H}}{{\partial {k_y}}}\left| {n{\bf{k}}} \right\rangle  - \left( {n \leftrightarrow m} \right)}}{{{{\left( {{E_{n{\bf{k}}}} - {E_{m{\bf{k}}}}} \right)}^2}}}} d{k_x}d{k_y}} ,
\label{ChernNumber}
\end{equation}
where $\Omega$ is the first Brillouin zone and we omit the ``eff" subscript of $H$.

We perform the Fourier transform of the Hamiltonian (\ref{EffectiveHamiltonian}) along the $x$ axis and impose vanishing boundary conditions on the $y$ axis. The resulting Hamiltonian for the SkX nanoribbon is\cite{OstahiePRB2018}
\begin{equation}
\begin{array}{l}
{H_{{\rm{eff}}}} = \sum\limits_{{k_x},m, < s,s' > } {t_{{\rm{eff}}}^{s,s'}d_{{k_x},m}^{s\dag }d_{{k_x},m}^{s'}}  + \sum\limits_{{k_x},m, \ll s,s' \gg } {t{'_{{\rm{eff}},s,s'}}d_{{k_x},m}^{s\dag }d_{{k_x},m}^{s'}} \\
 + \left[ {\sum\limits_{{k_x},m, < s,s' > } {t_{{\rm{eff}}}^{s,s'}d_{{k_x},m + 1}^{s\dag }d_{{k_x},m}^{s'}}  + \sum\limits_{{k_x},m, < s,s' > } {t_{{\rm{eff}}}^{s,s'}d_{{k_x},m}^{s\dag }d_{{k_x},m}^{s'}{e^{ - 2i{k_x}\lambda }}}  + {\rm{H}}{\rm{.c}}{\rm{.}}} \right]\\
 + \left[ {\sum\limits_{{k_x},m, \ll s,s' \gg } {t{'_{{\rm{eff}},s,s'}}d_{{k_x},m + 1}^{s\dag }d_{{k_x},m}^{s'}}  + \sum\limits_{{k_x},m, \ll s,s' \gg } {t{'_{{\rm{eff}},s,s'}}d_{{k_x},m}^{s\dag }d_{{k_x},m}^{s'}{e^{ - 2i{k_x}\lambda }}}  + {\rm{H}}{\rm{.c}}{\rm{.}}} \right].
\end{array}
\label{EffectiveNanoribbon}
\end{equation}

To investigate the topological properties of the SkX away from the strong-coupling limit, we directly work on the Hamiltonian (\ref{HInSublatticeManner}) with arbitrary magnitude of the Hund's coupling. To calculate the Chern number of the bulk bands, we perform 2D Fourier transform of Eq. (\ref{HInSublatticeManner}) and the resulting Hamiltonian reads
\begin{equation}
\begin{array}{l}
H = t\sum\limits_{{\bf{k}}, < s,s' > } {c_{\bf{k}}^{s\dag }c_{\bf{k}}^{s'}}  + t'\sum\limits_{{\bf{k}}, \ll s,s' \gg } {c_{\bf{k}}^{s\dag }c_{\bf{k}}^{s'}}  - J\sum\limits_{{\bf{k}},s} {{{\bf{n}}_s}c_{\bf{k}}^{s\dag }{\bf{\sigma }}c_{\bf{k}}^s} ,\\
 + \left[ {t\sum\limits_{{\bf{k}}, < s,s' > } {c_{\bf{k}}^{s\dag }c_{\bf{k}}^{s'}{e^{ - 2i{k_y}\lambda }}}  + t\sum\limits_{{\bf{k}}, < s,s' > } {c_{\bf{k}}^{s\dag }c_{\bf{k}}^{s'}{e^{ - 2i{k_x}\lambda }}}  + {\rm{H}}{\rm{.c}}{\rm{.}}} \right]\\
 + \left[ {t'\sum\limits_{{\bf{k}}, \ll s,s' \gg } {c_{\bf{k}}^{s\dag }c_{\bf{k}}^{s'}{e^{ - 2i{k_y}\lambda }}}  + t'\sum\limits_{{\bf{k}}, \ll s,s' \gg } {c_{\bf{k}}^{s\dag }c_{\bf{k}}^{s'}{e^{ - 2i{k_x}\lambda }}}  + {\rm{H}}{\rm{.c}}{\rm{.}}} \right].
\end{array}
\label{BulkHWithJ}
\end{equation}
This Hamiltonian is diagonal in the momentum apace and a $50 \times 50$ matrix in the ${\rm{sublattice}} \otimes {\rm{spin}}$ space. By solving Eq. (\ref{BulkHWithJ}), we can obtain the band structure and the electronic states of the SkX and the Chern number of the bands can be calculated from Eq. (\ref{ChernNumber}).
By performing Fourier transform along the $x$ axis and impose vanishing boundary conditions on the $y$ axis to Eq. (\ref{HInSublatticeManner}), we arrive at the Hamiltonian of the nanoribbon geometry of the SkX with arbitrary magnitude of the Hund's coupling
\begin{equation}
\begin{array}{l}
H = t\sum\limits_{{k_x},m, < s,s' > } {c_{{k_x},m}^{s\dag }c_{{k_x},m}^{s'}}  + t'\sum\limits_{{k_x},m, \ll s,s' \gg } {c_{{k_x},m}^{s\dag }c_{{k_x},m}^{s'}}  - J\sum\limits_{{k_x},m,s} {{{\bf{n}}_s}c_{{k_x},m}^{s\dag }{\bf{\sigma }}c_{{k_x},m}^s} ,\\
 + \left[ {t\sum\limits_{{k_x},m, < s,s' > } {c_{{k_x},m + 1}^{s\dag }c_{{k_x},m}^{s'}}  + t\sum\limits_{{k_x},m, < s,s' > } {c_{{k_x},m}^{s\dag }c_{{k_x},m}^{s'}{e^{ - 2i{k_x}\lambda }}}  + {\rm{H}}{\rm{.c}}{\rm{.}}} \right]\\
 + \left[ {t'\sum\limits_{{k_x},m, \ll s,s' \gg } {c_{{k_x},m + 1}^{s\dag }c_{{k_x},m}^{s'}}  + t'\sum\limits_{{k_x},m, \ll s,s' \gg } {c_{{k_x},m}^{s\dag }c_{{k_x},m}^{s'}{e^{ - 2i{k_x}\lambda }}}  + {\rm{H}}{\rm{.c}}{\rm{.}}} \right].
\end{array}
\label{NanoribbonHWithJ}
\end{equation}

\section{Results and discussions}

\subsection{Topological phase transition driven by polarity change}

We first consider the topological phase transition driven by the polarity change in the case of $t'=0$ and $J \gg t$. Using Eqs. (\ref{BulkHInKSpace}) to (\ref{EffectiveNanoribbon}), the topological properties of the electronic states for different polarity $Q_{\rm{sk}}$ are shown in Figs. 2 to 3. For the polarity value $Q_{\rm{sk}}=1$, which is also the topological number of the skyrmion, the spin field demonstrates a monopole profile. For the polarity value $Q_{\rm{sk}}=2$, the SkX has a high topological number and the spin field demonstrates a dipole profile. While the polarity of the skyrmion changes continuously from the monopole to the dipole, the Chern number of each band retains quantized of unity up to $Q_{sk}=1.3$, and then the Chern number of the bands bears indefinite values from $Q_{\rm{sk}}=1.3$ to $Q_{\rm{sk}}=1.7$, and finally the Chern number of the bands retains to be 0, 1, 0, 1, $\cdots$, consecutively from $Q_{\rm{sk}}=1.7$ to $Q_{\rm{sk}}=2$ (See Fig. 3). This phenomenon demonstrates a topological phase transition from the monopole lattice phase to the dipole lattice phase. To confirm the topological properties of each phase, we plotted the electronic spectrums of the SkX nanoribbon for $Q_{\rm{sk}}=1$, 1.2, 1.5, 1.8, and 2 in Fig. 2 (k)-(h), respectively. The edge states within the bulk band gaps are clearly seen. For $Q_{\rm{sk}}=1$ and 1.2, the number of the crossing of the edge states is 1, 3, 5, and 7, consecutively. The crossing in the second band gap at the two Brillouin zone boundaries is counted once. The local density of states of the edge states close to one of the crossing knot is shown in Fig. 4. It is shown that these two states strictly localized at the outermost edge skyrmion unit cell and edge site, which is a strong signal of the topological edge state. For $Q_{\rm{sk}}=1.8$ and 2, the number of the crossing of the edge states is 1, 2, 1, and 4, consecutively. This topological phase is a previously unknown. For the edge states in the first and third band gap above a band with the Chern number equal to 0 and below a band with the Chern number equal to 1, the number of the crossing knots can be calculated from $2 \times 0 +1$. For the edge states in the second band gap above a band with the Chern number equal to 1 and below a band with the Chern number equal to 0, the number of the crossing knots can be calculated from $2 \times 1 +0$. For the edge states in the fourth band gap above two bands with the Chern number equal to 1 and one band with the Chern number equal to 0 and below a band with the Chern number equal to 0, the number of the crossing knots can be calculated from $2 \times 2 +0$. This interpreted the bulk-edge correspondence of the dipole lattice phase. For $Q_{\rm{sk}}=1.5$, the Chern number of the bands are 0, 2, 1, and 1, consecutively (see Fig. 3). From this we can see that it is neither a monopole lattice phase nor a dipole lattice phase. The Chern number of the bands varies while $Q_{\rm{sk}}$ increases continuously from 1.3 to 1.7. Because the phase does not have a definite Chern number or a definite Chern number series, which could span a finite region in the parameter space, we call it an indefinite phase. Also, this phase is not topological and does not have a bulk-edge correspondence. From Fig. 2 (m), we could see that the number of the crossing knots of the edge states are 1, 3, 4, and 7, which could not be calculated from the bulk band Chern number by topological rules.

\subsection{Topological phase transition driven by next-nearest-neighbor hopping}

In this section, we consider the topological phase transition driven by the nnn hopping in the case of $Q_{\rm{sk}}=1$ and $J \gg t$. Using Eqs. (\ref{BulkHInKSpace}) to (\ref{EffectiveNanoribbon}), the topological properties of the electronic states for different nnn hopping energy $t'$ are shown in Figs. 5 and 6. From Fig. 5, a clear phase transition from the monopole lattice with the Chern number of each band equal to 1 to an indefinite phase can be seen to occur at $t' \approx 0.47 t$. To confirm the topological properties of the monopole lattice phase with the nnn hopping present, we plotted the electronic spectrums of the SkX nanoribbon for $t'=0.2 t$ and $t'=0.4 t$ in Fig. 6 (e) and (h), respectively. The edge states within the bulk band gaps are clearly seen. Due to the effect of the nnn hopping, for $t' < 0.47 t$, the bulk bands are raffled relative to the case of $t' =0$, but still directly gapped, without changing the topological properties of the electronic states. As shown in the figure, the number of the crossing of the edge states is 1, 3, 5, and 7, consecutively, in the monopole lattice phase. The crossings at the two Brillouin zone boundaries with identical eigenenergy values are counted once. For $t' > 0.47 t$, the electronic bands are stacked together with the direct band gaps closed. This is because the nnn hopping changes the flux pattern of the emergent magnetic field. For large nnn hopping strengths, the topological order is broken giving rise to an indefinite nontopological phase, as shown in the right two columns of Fig. 6. We could also see that the energy levels are lifted by the nnn hopping because we assumed $t'$ to be in the same sign with $t$ as expressed in Eq (\ref{Hamiltonian})\cite{HamamotoPRB2015}.

\subsection{Effect of finite Hund's coupling on the topological phase transitions }

In the previous two subsections, we considered the topological phase transitions of the electronic states in the SkX driven by the polarity change and the nnn hopping in the strong coupling limit. In this subsection, we consider the effect of finite Hund's coupling. Using Eqs. (\ref{BulkHWithJ}) and (\ref{NanoribbonHWithJ}), we calculated the Chern number of the bulk bands and the edge states of the nanoribbon of the electronic states in the SkX. Different from the effective model (\ref{EffectiveHamiltonian}) of spinless electrons, the electrons in the original double exchange model (\ref{Hamiltonian}) have a free spin, which doubles the dimension of the Hilbert space of the electrons into the ${\rm{momentum}} \otimes {\rm{sublattice}} \otimes {\rm{spin}}$. The Hamiltonian is diagonal in the momentum space and a $50 \times 50$ matrix in the ${\rm{sublattice}} \otimes {\rm{spin}}$ space. As a result, the number of bulk bands doubles. So we calculated the Chern number of the bulk and the edge states of the nanoribbon for the lowest eight bands. Numerical results are presented in Figs. 7-9.

Fig. 7 plots the effect of finite $J$ on the monopole lattice phase with $Q_{\rm{sk}}=1.2$ and $t' =0$. From the figure we can see that with decreasing $J$ from $5 t$ to about $1.4 t$ the Chern numbers of the eight bands are precisely equal to 1, which is consistent with the results of the case of infinite $J$ shown in Fig. 3. When $J$ is smaller than $1.4 t$, the Chern numbers deviate from quantized values, indicating an indefinite nontopological phase. This can be interpreted by that when the strength of the Hund's coupling becomes comparable or smaller than the strength of nn hopping, the spin of the conducting electrons no longer completely align with the local magnetization, the topology of the electrons gained from the spin field is broken, and the flux of the emergent magnetic field becomes irregular breaking the quantization of the Hall conductivity. On the other hand, the $J \approx 1.4 t$ crossover demonstrates that the topological properties obtained from the effective model holds well to relatively small coupling strength.

Fig. 8 plots the effect of finite $J$ on the monopole lattice phase with the nnn hopping present with $Q_{\rm{sk}}=1$ and $t' =0$. From the figure we can see that with decreasing $J$ from $2 t$ to about $0.9 t$ the Chern numbers of the eight bands are precisely equal to 1, which is consistent with the results of the case of infinite $J$ shown in Fig. 5. When $J$ is smaller than $0.9 t$, the Chern numbers deviate from quantized values, indicating an indefinite nontopological phase. This crossover value is smaller than that shown in Fig. 7. There are two reasons. One is that Fig. 7 plots the result of $Q_{\rm{sk}}=1.2$. This corresponds to an imperfect monopole spin field as shown in Fig. 2 (b). Although the topological phase tolerate this deviation in the strong coupling limit, it needs larger coupling strength to hold the topological phase in the case of $Q_{\rm{sk}}=1.2$ than in the case of $Q_{\rm{sk}}=1$. the other reason is that when the nnn hopping is present, the traversing pattern of the flux of the emergent magnetic field is changed. This change may favor the topological phase in particular parameter settings. When the topological phase of the strong coupling limit is broken at $J \approx 0.9 t$, we see that the effective model obtained from the strong coupling approximation works well to realistic finite coupling strength.

In Fig. 9, we show the electronic spectrum of the bulk and the edge states of the nanoribbon with finite $J$. Fig. 9 (a) and (c) are the finite-$J$ counterpart of Fig. 2 (g) and (l). Fig. 9 (b) and (d) are the finite-$J$ counterpart of Fig. 6 (a) and (e). Similar to the previous figures, in the case of $J=2t$, $Q_{\rm{sk}}=1.2$, and $t'=0$, the bulk spectrums are slightly tilted as a result of the tilted spin field, but the edge states retain the topological properties with the number of the crossing of the edge states equal to 1, 3, 5, 7, $\cdots$ consecutively; in the case of $J=1.5 t$, $Q_{\rm{sk}}=1$, and $t'=0.2 t$, the bulk spectrums have more hump-dip patterns as a result of the nnn hopping, but the edge states retain the topological properties with the number of the crossing of the edge states following the topological rule of the bulk-edge correspondence.

\section{Conclusions}

In this work, we extend the double exchange model by taking into account the nnn hopping. Techniques to solve the Hamiltonian within and without the strong coupling limit are elaborated. Numerical results of the Chern number of the bands of the bulk SkX as well as the edge states of the nanoribbon demonstrate a phase transition from the monopole lattice to the dipole lattice through an intermediate indefinite phase driven by the polarity change. When the strength of the nnn hopping increases from zero, the topological phase is broken at $t' \approx 0.47 t$ for the monopole SkX with $Q_{\rm{sk}}=1$. When the Hund's coupling is reduced from the infinite limit to comparable or smaller than the nn hopping energy, the topological phase is broken as a result of the release of the spin degree of freedom of the conducting electrons.

The present work demonstrates analogy between the electronic states in the SkX and those in Dirac-Weyl semimetals in consideration of the topological transport properties such as the quantized Hall conductivity\cite{HamamotoPRB2015} and topological phase transition driven by the nnn hopping\cite{BeugelingPRB2012}. Following this theme line, we could foresee further investigations of other transport properties such as tunneling, proximity-induced superconducting\cite{KubetzkaPRMat2020}, and etc., which effects are extremely sensitive to the nontrivial topology of the conducting electronic bands as seen in various types of Dirac-Weyl materials.

The experimental relevance of the present work lies in two aspects. One is that the real SkX can have polarity deviation from the cases of $Q_{\rm{sk}}$ exactly equal to 1 or two because numerical simulation demonstrates continuous change of $Q_{\rm{sk}}$ of isolated skyrmions in various situations\cite{XZhangSR2015, XiangjunXingPRB2016}. Also the nnn hopping in the SkX with atomic square lattice\cite{MuhlbauerScience2009} may not be neglected. The other aspect is that recent numerical results predict the skyrmion phase made up of superfluid cold atoms, which lends one the hope of simulating the dynamics of the SkX in the cold atom systems in the future\cite{XiaoLongChenPRRes2020}.

\section{Acknowledgements}

We acknowledge support by the National Natural Science Foundation of China (No. 11004063) and the Fundamental Research Funds for the Central Universities, SCUT (No. 2017ZD099).

\clearpage

\begin{figure}[ht]
\includegraphics[height=12cm, width=14cm]{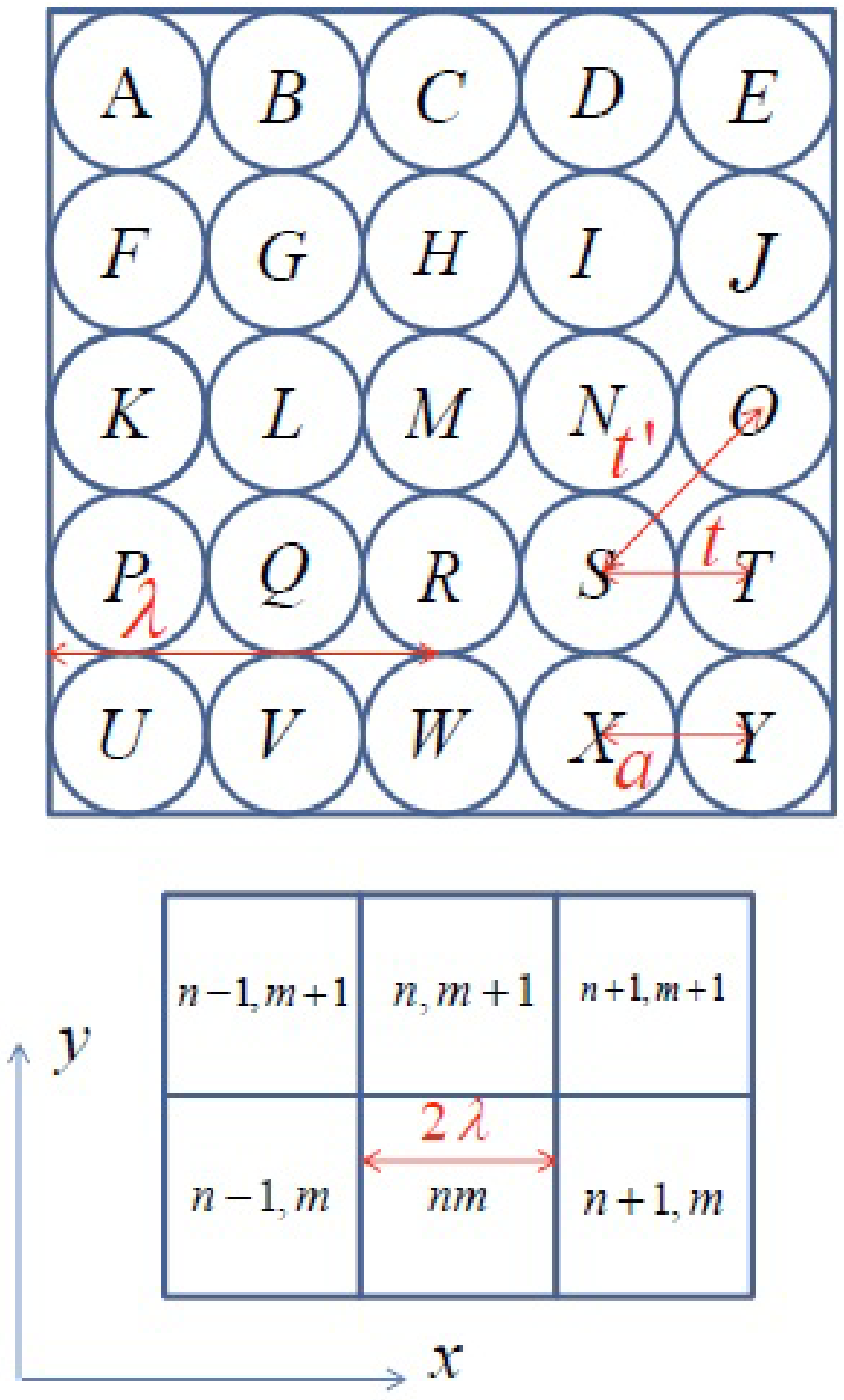}
\caption{ The SkX has two lattice structures. One is the atomic lattice. The other is the lattice of the skyrmion with each skyrmion constituting a giant unit cell. We consider both of them to be square lattices. In this figure, the upper panel is the sublattice structure within each skyrmion unit cell and the lower panel is the skyrmion lattice structure. The 2D SkX extends in the $x$-$y$ plane with each skyrmion unit cell containing $5 \times 5 =25$ atoms. $a$ is the atomic lattice constant. $\lambda$ is the radius of the skyrmion vortex, which is shown in both panels. $t$ and $t'$ indicate nn and nnn hopping, respectively. Capital letters A to Y labels different sublattices in the unit cell. $n$ and $m$ are the coordinates of the skyrmion unit cell in the $x$ and $y$ direction, respectively.
 }
\end{figure}
\clearpage
\begin{figure}[ht]
\includegraphics[height=12cm, width=15cm]{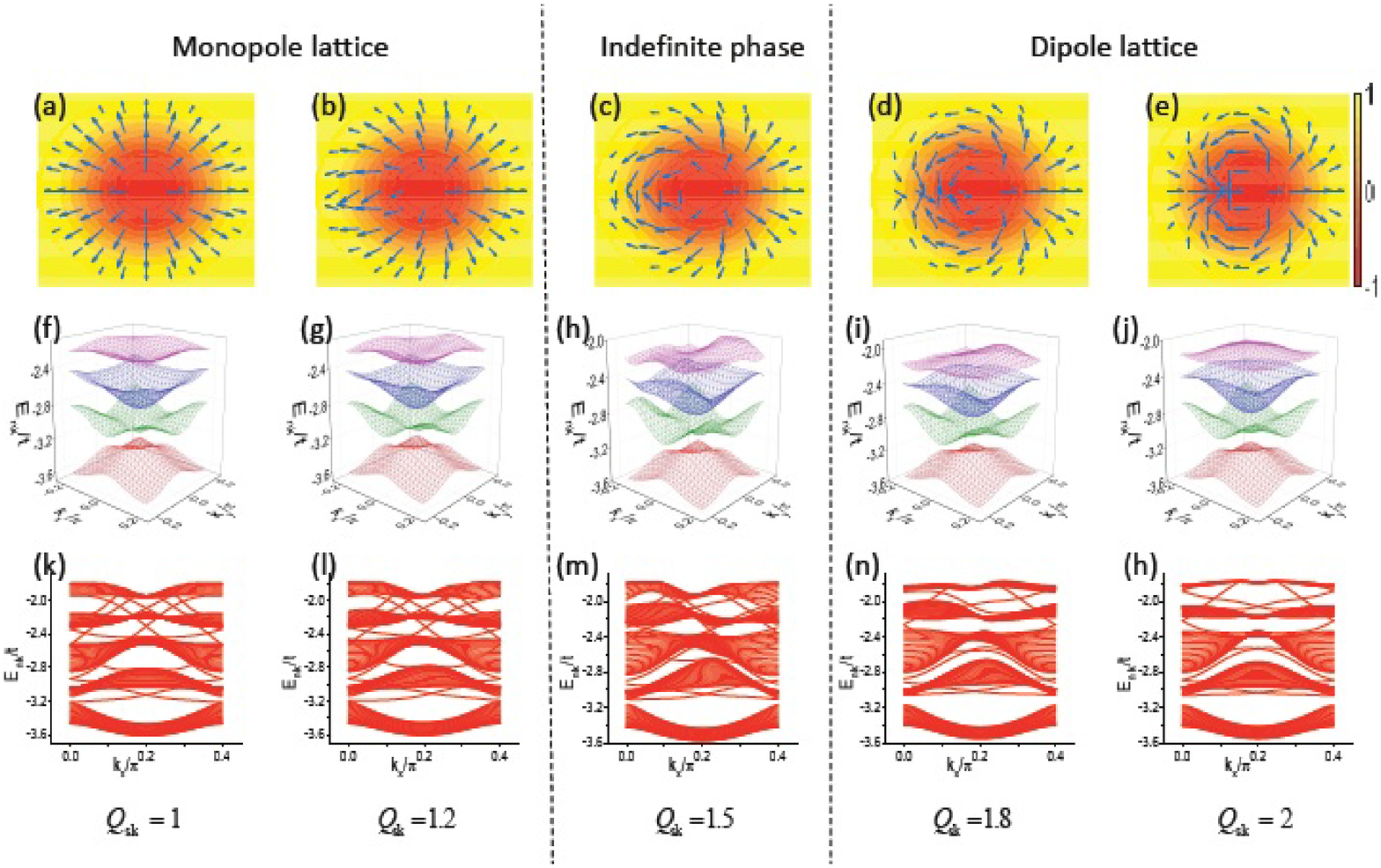}
\caption{Skyrmion profile (a)-(e), bulk electronic bands (f)-(j), and spectrum of nanoribbon (k)-(h) in the SkX with different values of $Q_{\rm{sk}}$. In (a)-(e), the arrows and background color are the $x$-$y$ and $z$ components of the spin field of the skyrmion, respectively. For $Q_{\rm{sk}}=1$ and $1.2$ in the left two columns, the SkX is in the monopole lattice phase with the $x$-$y$ component of the spin field of each skyrmion demonstrating a close-to monopole profile, the Chern number of each band equal to 1, and the number of the edge-state crossing within the bulk band gaps equal to 1, 3, 5, and 7, consecutively. For $Q_{\rm{sk}}=1.8$ and $2$ in the right two columns, the SkX is in the dipole lattice phase with the $x$-$y$ component of the spin field of each skyrmion demonstrating a close-to dipole profile, the Chern number of the bands equal to 0, 1, 0, and 1, consecutively and the number of edge-state crossing within the bulk band gaps equal to 1, 2, 1, and 4 consecutively. This figure plots the case of $t'=0$ in the strong coupling limit.  }
\end{figure}
\clearpage

\begin{figure}[ht]
\includegraphics[height=12cm, width=15cm]{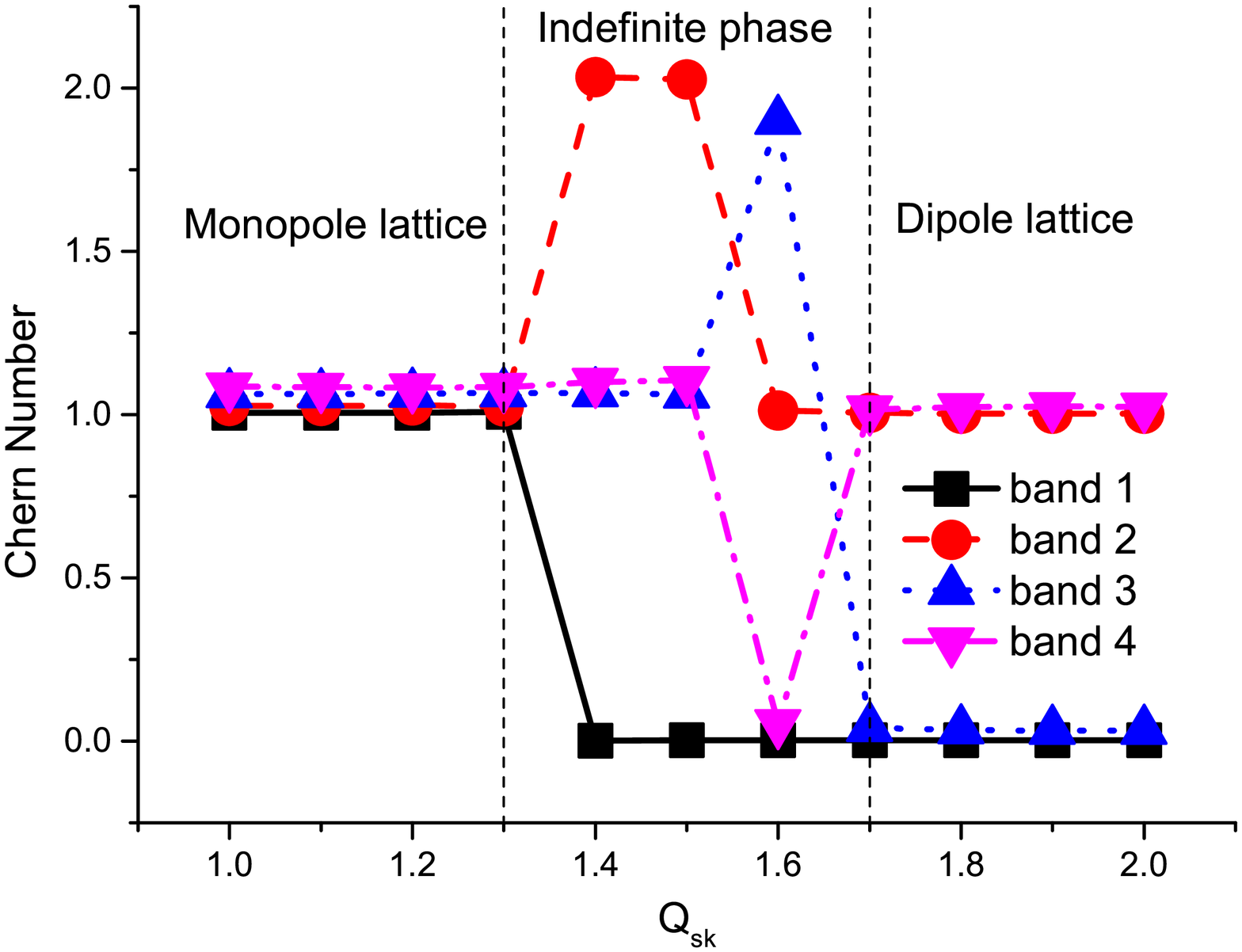}
\caption{Chern number of the lowest four bands as a function of $Q_{\rm{sk}}$. This figure plots the case of $t'=0$ in the strong coupling limit. Phase boundaries are indicated by vertical dashed lines. The phase transition between the monopole lattice phase and the indefinite phase occurs at $Q_{\rm{sk}} \approx 1.3$. The phase transition between the indefinite phase and the dipole lattice phase occurs at $Q_{\rm{sk}} \approx 1.7$.  }
\end{figure}
\clearpage

\begin{figure}[ht]
\includegraphics[height=12cm, width=15cm]{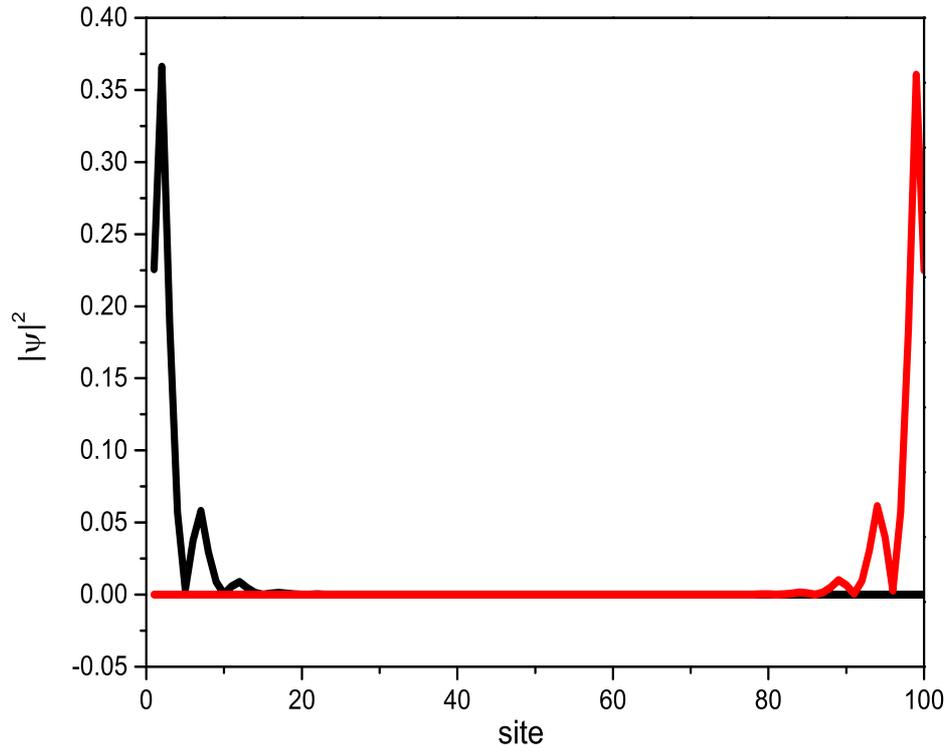}
\caption{ The local density of states of the edge states in a nanoribbon geometry. The two states are picked close to the lowest spectrum crossing in Fig. 2 (k) with $k_x =0.49 {\pi} / {\lambda} \approx 0.2 \pi$. The sites are along the direction perpendicular to the nanoribbon direction, which is $y$ direction in Eq. (\ref{EffectiveNanoribbon}). }
\end{figure}
\clearpage

\begin{figure}[ht]
\includegraphics[height=12cm, width=17cm]{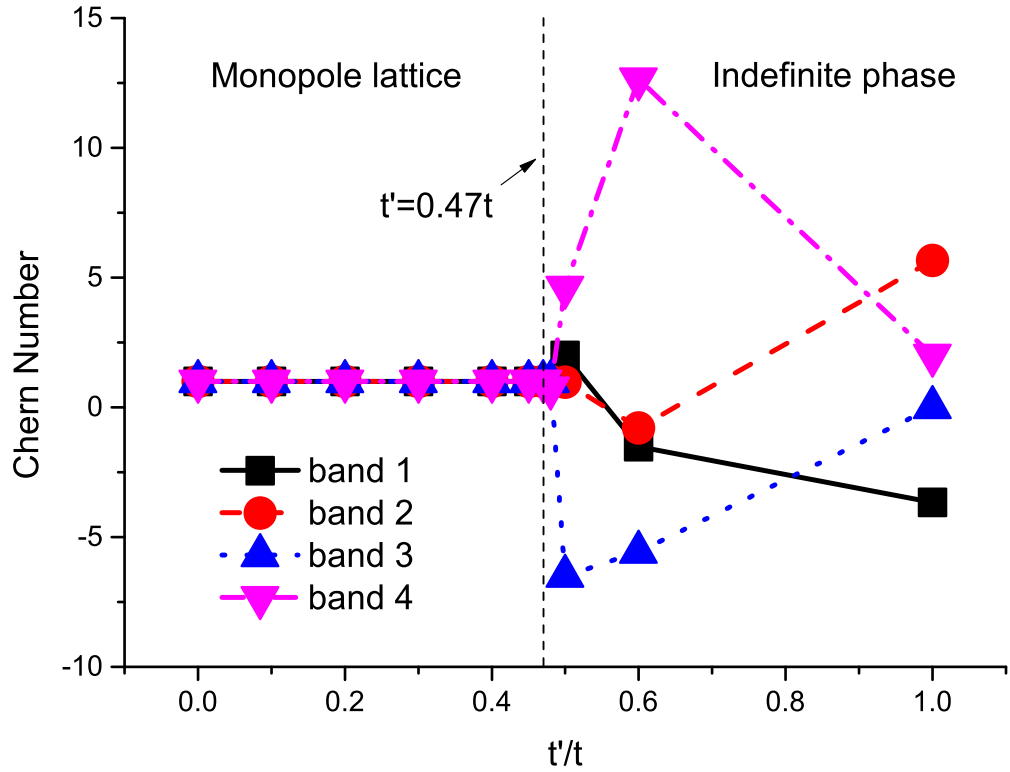}
\caption{Chern number of the lowest four bands as a function of $t'$. This figure plots the case of $Q_{\rm{sk}}=1$ in the strong coupling limit. The phase transition between the monopole lattice phase and the indefinite phase occurs at $t' \approx 0.47 t$, which is indicated by a vertical dashed line. }
\end{figure}
\clearpage

\begin{figure}[ht]
\includegraphics[height=12cm, width=15cm]{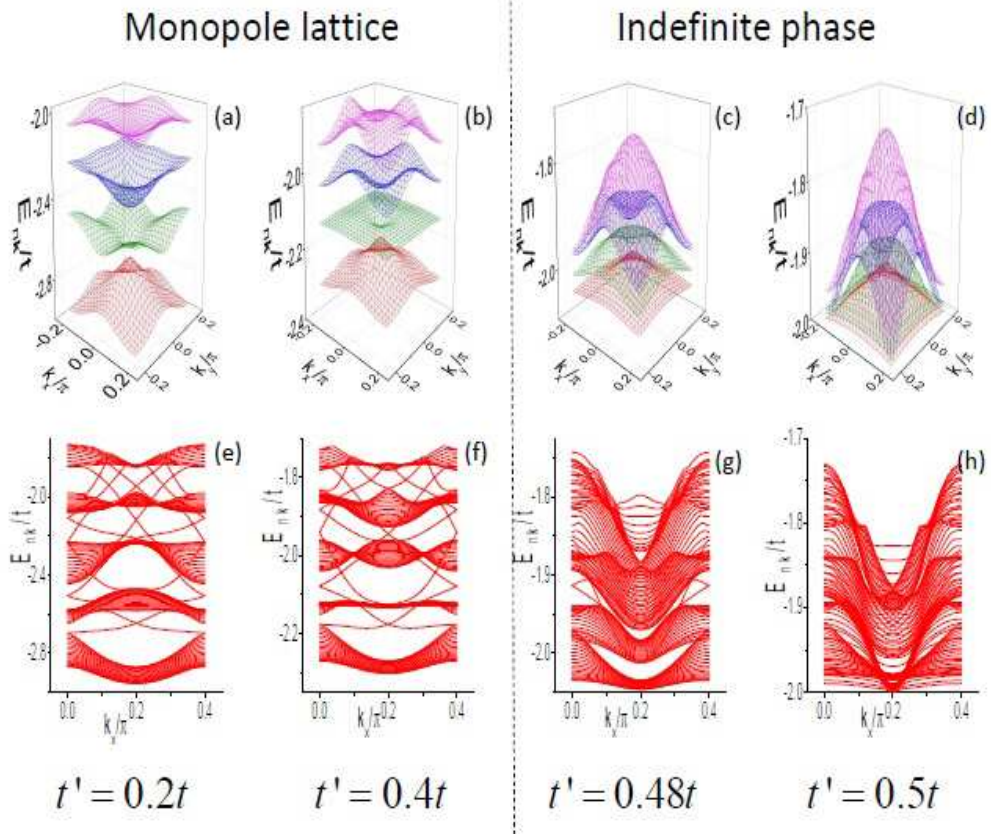}
\caption{Bulk electronic bands (a)-(d) and spectrum of nanoribbon (e)-(h) in the SkX with different values of $t'$. For $t' =0.2 t$ and $0.4 t$ in the left two columns, the SkX is in the monopole lattice phase, the Chern number of each band is equal to 1, and the number of the edge-state crossing within the bulk band gaps equal to 1, 3, 5, and 7, consecutively. For $t' =0.48 t$ and $0.5 t$ in the right two columns, the SkX is in an indefinite phase with indefinite Chern number of the bulk bands. This figure plots the case of $Q_{\rm{sk}}=1$ in the strong coupling limit.  }
\end{figure}
\clearpage

\begin{figure}[ht]
\includegraphics[height=12cm, width=15cm]{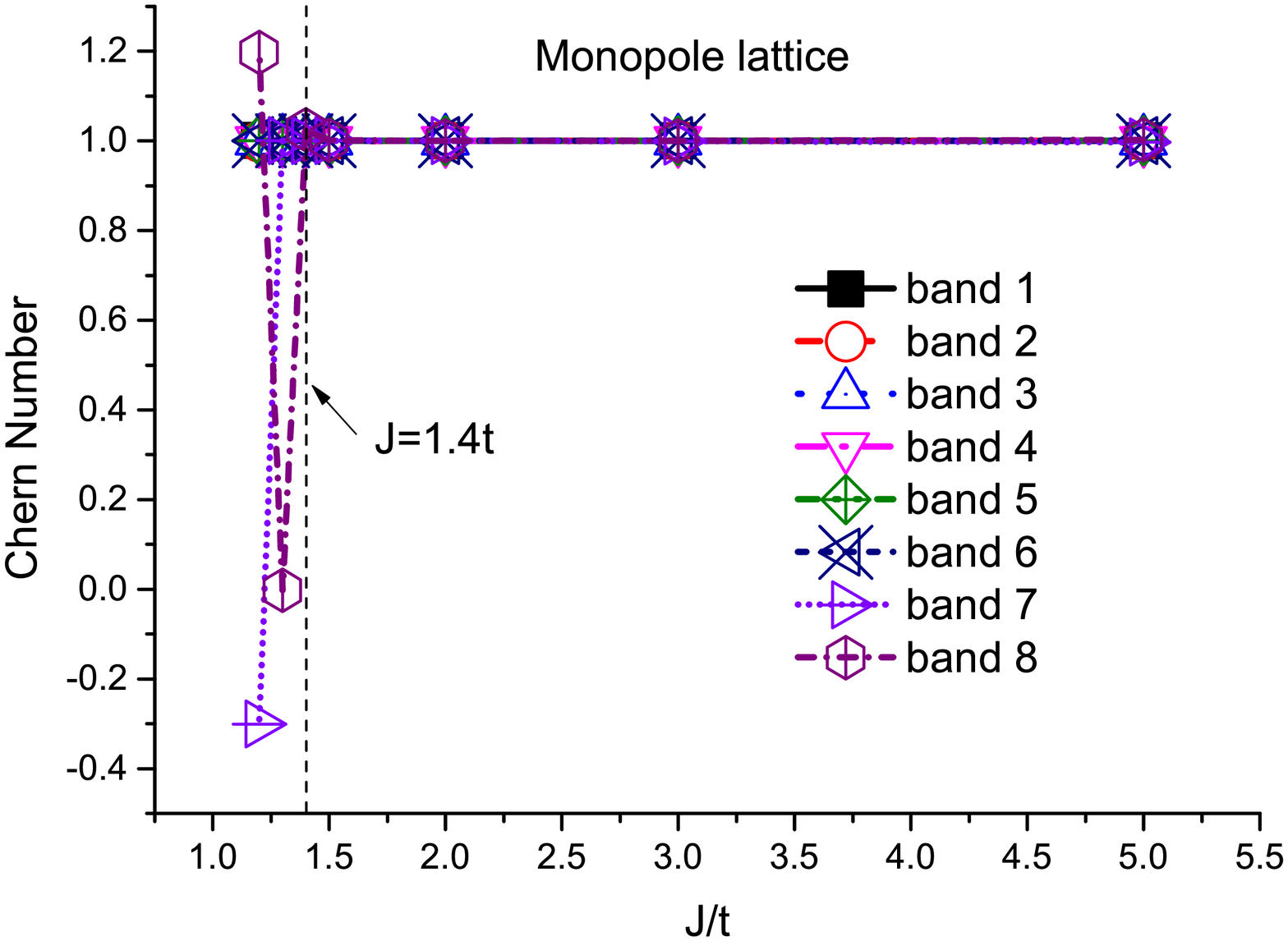}
\caption{Chern number of the lowest eight bands as a function of $J$. This figure plots the effect of $J$ on the monopole lattice phase with $Q_{\rm{sk}}=1.2$ and $t' =0$. The deviation from the monopole lattice phase occurs at $J \approx 1.4 t$, which is indicated by a vertical dashed line.}
\end{figure}
\clearpage

\begin{figure}[ht]
\includegraphics[height=12cm, width=15cm]{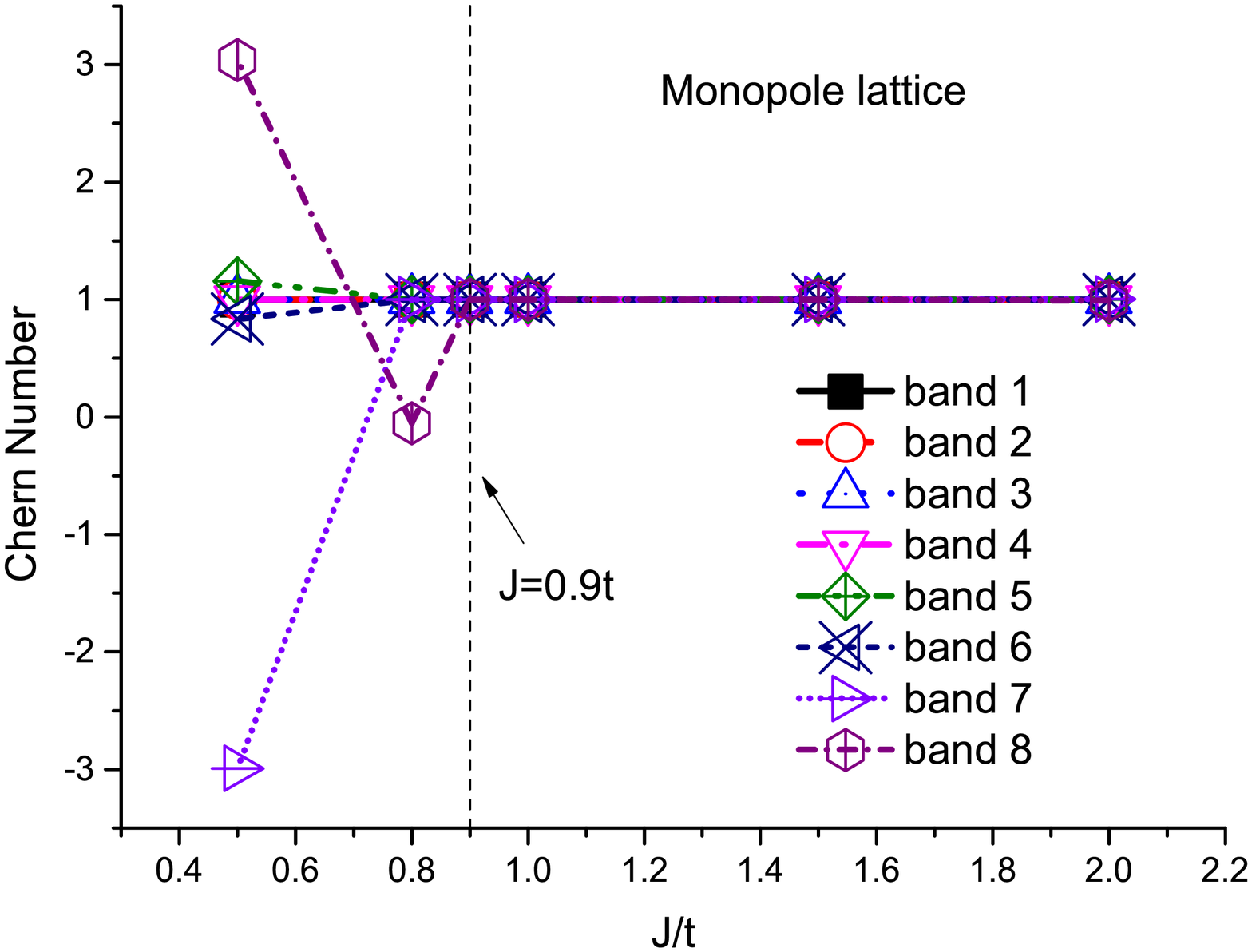}
\caption{Chern number of the lowest eight bands as a function of $J$. This figure plots the effect of $J$ on the monopole lattice phase with $Q_{\rm{sk}}=1$ and $t' =0.2 t$. The deviation from the monopole lattice phase occurs at $J \approx 0.9 t$, which is indicated by a vertical dashed line. }
\end{figure}
\clearpage

\begin{figure}[ht]
\includegraphics[height=12cm, width=15cm]{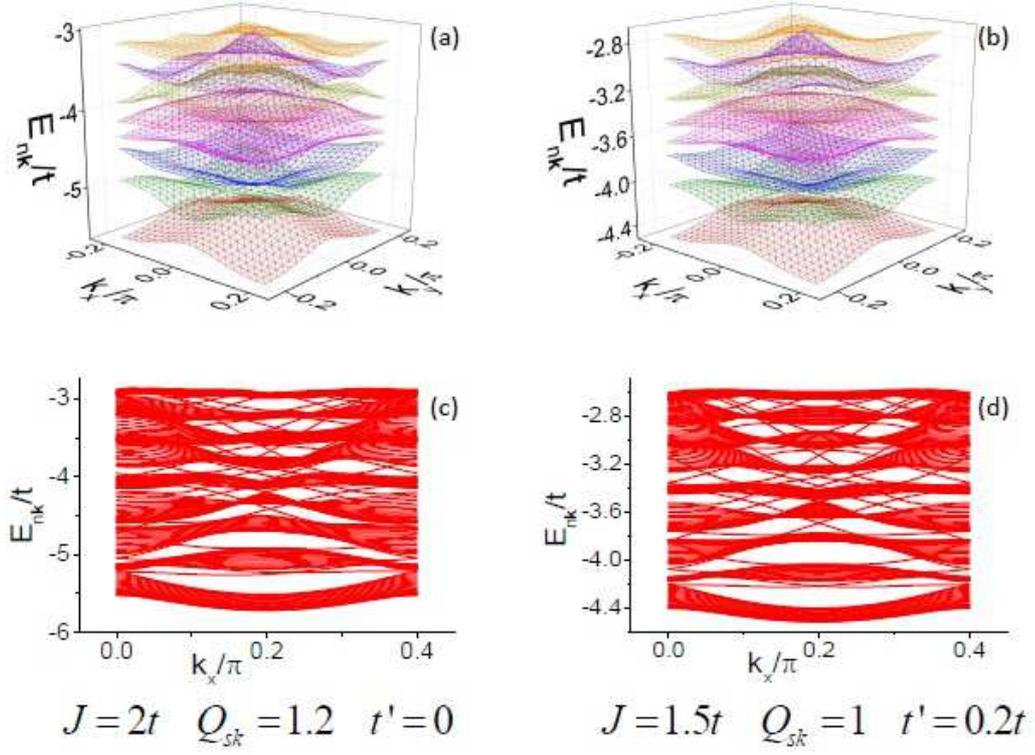}
\caption{Bulk electronic bands (a)(b) and spectrum of nanoribbon (c)(d) in the SkX with different values of $J$, $Q_{\rm{sk}}$, and $t'$. For all the subfigures, the SkX is in the monopole lattice phase, the Chern number of each band is equal to 1, and the number of the edge-state crossing within the bulk band gaps equal to 1, 3, 5, 7, $\cdots$ consecutively.   }
\end{figure}
\clearpage

\end{document}